\documentstyle{article}

\begin{document}
\title{Metastable superpositions of ortho- and para-Helium states}
\author{ P. Sancho (a), L. Plaja (b)  \\ (a) GPV de
Valladolid. Centro Zonal
en Castilla y Le\'on. \\ AEMET. Ori\'on 1, 47014, Valladolid, Spain \\
(b) Area de Optica. Departamento de F\'{\i}sica Aplicada.
Universidad
\\ de Salamanca. Pl. de la Merced s/n, 37008, Salamanca, Spain }
\date{}
\maketitle
\begin{abstract}
We analyze superpositions of ortho- and para-Helium states,
considering the possible existence of stationary and metastable
states in the system. In particular, the metastable superposition of
$1s2s$ ortho and para states seems to be accessible to experimental
scrutiny.
\end{abstract}

\section{Introduction}
One of the most interesting manifestations of the principle of
antisymmetrization of two-fermion systems is the existence of two
types of configurations for the Helium atom. They correspond to
singlet and triplet states and are usually denoted as para- and
ortho-Helium states, showing different distributions of energy
levels.

The study of the Helium atom is not a closed subject \cite{Tan}. For
instance, calculations of some energy levels of the confined Helium
atom \cite{Bar} and refinements in the para- and ortho-Helium evaluations
\cite{Dua} have been presented in the literature. In has also been
shown that the ionization properties of the Helium atom are strongly
dependent on the type of configuration \cite{Pla}. The antiprotonic
Helium \cite{Ead}, the system produced by the capture of an
antiproton by a $He^+$ ion, has also extensively been studied.

In this Letter we suggest a new line to study the interesting
properties of para- and ortho-Helium. We shall show that it is
possible to prepare the Helium atom in a metastable superposition of
para and ortho states. The scheme of preparation is in principle
very simple. It is based on the capture by an $He^+$ ion of an
electron, which must be prepared in a superposition of spin states.
As we shall see later, in order to obtain a superposition one must
only demand to the Hamiltonian describing the capture process to be
symmetric. We should stress that this scheme appears quite naturally
during the double ionization of He with intense laser fields. It has
been shown that, for laser intensities below $5 \times 10^{15}$
W/cm$^2$, the relevant path for ionization is a non-sequential
process, in which the two electrons entangled are emitted with the
same direction \cite{walker}. Since the probability for single
ionization is always higher than the double, there is also a
fraction of $He^+$ so the electron capture is also feasible. Note
however that this process involves three particles, therefore the
final superposition of He states is entangled with the surviving
electron. Up to our knowledge, superpositions of ortho and para
states have only been considered in the context of Helium collisions
\cite{JPB} (see section 3).

The proposed states would be interesting in several aspects. From a
fundamental point of view they would provide us with a situation
where the linearity of quantum mechanics has not been explored
before, that in which the antisymmetrization postulate and exchange
effects must be taken into account. Superpositions of different
energy states of an atom have previously been considered in the
literature. An atom can be prepared in a superposition of excited
energy states using an impulsive excitation such that its Fourier
spectrum contains frequency components corresponding to the energy
intervals between ground and excited states. The rate of spontaneous
emission of atoms prepared in this manner can oscillate in time.
These quantum beats differ from the usual exponential decay expected
for atoms in well-defined energy states \cite{Sil}. Quantum beats,
manifested as modulations in the absorption rates, are also present
in the absorption of light by atoms in superposition states
\cite{Wol}. However, in these (energy-type) superpositions the
exclusion principle does not play any role.

From a more practical point of view, it has been discussed the
existence of collisional velocity changes associated with atoms in
superposition states \cite{Mos}. It has also been signaled that
atomic superposition states are sensitive to phase dependent
properties of radiation fields and, consequently, could be employed
as detectors \cite{Kni}. Moreover, we can expect that many
properties of the atom such as absorption rates, ionization energy ,
etc. will differ in normal and superposition states and could be
useful to understand the mechanisms involved in these processes.

Before considering these possibilities we must analyze the stability
of the states of the superposition. In general, as we shall see in
Sect. 3, there are only stationary states in some particular cases,
when the ortho and para states are degenerate or almost degenerate.
In the absence of stationary states we must consider the existence
of metastable states, which could be studied with high resolution
laser spectroscopy techniques. We shall study the superposition of
ortho- and para-1s2s states, showing that it can be prepared as a
long lifetime metastable superposition state. Moreover, in
principle, this state is accessible to interesting experimental
verifications such as the modification of the amplitude of the
quantum beats and the variation of the mean lifetime and of the
fluctuations of the decaying rate. In this context it must be
signaled that the metastable $1s2s$ ortho-Helium state has been Bose
condensed \cite{Sci}.

\section{Preparation of the superposition}

First of all we show how to prepare the superposition. On the one
hand, we must have an $He^+$ ion with the electron in a well-defined
state of the spin component along a given axis, for instance
$|\uparrow >_z$. We take as axis of reference the $z$ one, denoting
by $|\uparrow >_z$ and $|\downarrow >_z$ the two possible states. We
can determine that the electron is in the correct state by a direct
measurement or by preparation. In the last case we should have an
$He^{++}$ ion, which captures an electron in the state $|\uparrow
>_z$.

On the other hand, once prepared the $He^+$ ion with the electron in
the $|\uparrow >_z$ state, we must have an electron in a
superposition state $|\phi >= \alpha |\uparrow >_z + \beta
|\downarrow >_z$, with $|\alpha |^2 +|\beta |^2=1$. This step can be
easily done by preparing the electron in a well-defined state along
other spin axis. For instance, the up and down components of the
spin along the two axes orthogonal to the $z$ one are $(|\uparrow
>_z + |\downarrow >_z)/\sqrt{2}$ and $(|\uparrow >_z -i |\downarrow
>_z)/\sqrt{2}$. An alternative procedure, as discussed above, is to
consider the non-sequential double ionization of He.
Now we can make interact the electron in state $|\phi >$ with the
$He^+$ ion, whose state is described by the ket $|He^+ >$.
If $He^+$ would only interact with an electron in state $|\uparrow
>_z$, capturing it, the map describing the interaction
would be
\begin{equation}
|He^+> |\uparrow >_z \rightarrow |He_{or}>
\end{equation}
We would obtain an ortho-Helium atom because we have assumed that
the other electron was also in the state $|\uparrow >_z$.
On the other hand, if the ion $He^+$ would interact with an electron
in the state $|\downarrow >_z$, the evolution would be
\begin{equation}
|He^+> |\downarrow >_z \rightarrow |He_{pa}>
\end{equation}
In this case we would obtain a para-Helium state.
Finally we move to the most interesting situation, that with the
incident electron in a superposition of spin states. If the capture
of the electron by the ion is a linear process (we discuss this
point later) we have
\begin{equation}
|He^+> |\phi > \rightarrow \alpha |He_{or}> + \beta |He_{pa}>
\end{equation}
which is a superposition of the Helium atom in para and ortho states.

We note that there is no superselection rule preventing the
superpositions considered here. From all the
superselection rules presented so far in the literature the only one
that is related to our proposal is that preventing the existence of
superpositions with different values of $(-1)^{2J}$, representing
$J$ the modulus of the total angular momentum, ${\bf J}={\bf L}+{\bf S}$, with
${\bf L}$ the orbital angular momentum and ${\bf S}$ the spin.
The two states of the superposition must have the same value of
$(-1)^{2J}$. In other words, $2J$ must be in both cases an even or
an odd integer. This is so in our case because the values of $l$ are
$0,1,2...$ and $s=1/2$. Then $2J=2l+1$ that is always odd.

We discuss now the linearity of the capture process. The Hamiltonian
describing the capture must take into account two different types of
interactions: (i)The electromagnetic, spin-orbit, spin-spin, etc.
usual interactions in the atom. (ii)The exchange effects associated
with the antisymmetrization of the wavefunction of two identical
electrons. With respect to (i) it is well-known the linearity of the
associated Hamiltonian. We consider now (ii). The wavefunction of
the complete system (nucleus plus the two electrons) must be
antisymmetrized with respect to the variables of the two electrons,
$\Psi ({\bf x},s_x;{\bf y},s_y;{\bf Z};t)= \psi ({\bf x},s_x;{\bf
y},s_y;{\bf Z};t) - \psi ({\bf y},s_y;{\bf x},s_x;{\bf Z};t)$ where
${\bf x}$ and ${\bf y}$ are the spatial coordinates of the two
identical particles, $s_x$ and $s_y$ refer to the spin components
and ${\bf Z}$ includes all the variables related to the nucleus. It
is simple to show adding the Schr\"{o}dinger equations ruling the
evolution of each $\psi $ that $\Psi$ only obeys a linear
Schr\"odinger's equation when $\hat{H} ({\bf x},s_x;{\bf y},s_y;{\bf
Z};t)=\hat{H} ({\bf y},s_y;{\bf x},s_x;{\bf Z};t)$ ($\hat{H}$ is the
Hamiltonian of the system), that is, when the Hamiltonian is
symmetric. Since all the Hamiltonians used in atomic physics fulfill
this condition we must expect the capture process to be linear.

Finally, we briefly discuss the possibility of actually implementing
the scheme suggested here. We start with a sample of $He$ atoms,
which is illuminated by a laser tuned in the adequate frequency to
induce double ionization. Using electric fields (which do not modify
the spins) we can separate the $He^{++}$ ions from $He$ atoms and
$He^+$ ions. Then a beam of electrons in the $|\uparrow >_z$ state
interacts with the sample of $He^{++}$ ions. The $He^+(|\uparrow
>_z)$ ions produced by the capture of one electron are separated,
using again an electric field, from the $He^{++}$ ions and
$He(|\uparrow >_z, |\uparrow >_z  )$ atoms. Finally, a beam of
electrons, for instance in the $|\uparrow >_x$ state, interacts with
the sample of $He^+(|\uparrow >_z)$ ions. The ions that capture
electrons become in a superposition state. Using once more an
electric field we can separate them from the $He^+(|\uparrow >_z)$
ions. The beams of electrons can be obtained from sources producing
them in arbitrary random spin states using Stern-Gerlach devices
with adequate orientations. An interesting alternative is provided
by the process of non-sequential double ionization of He in strong
electromagnetic process. In this case, the initial state is an
entangled electron pair ionized from some neighbor atom. The capture
of one of the electrons by a $He^+$ ion, leads to a entangled state
of the Helium atom with the surviving electron. The properties of
such three particle system will be the subject of a future
investigation.

\section{Stationary states of the superposition}

A fundamental question to be answered about the superposition is its
stability. In the quantum realm an atom can be stable because of the
existence of stationary states. We must look for the stationary
states of the superposition, which would be given by the solutions
of the equation
\begin{equation}
\hat{H} (\alpha |He_{or}[n]> + \beta |He _{pa}[m]>)=E_{[n,m]}
(\alpha |He_{or}[n]> + \beta |He _{pa}[m]>)
\end{equation}
where $[n]$ and $[m]$ represent the two sets of indexes
characterizing the stationary states of both types of
configurations. As $\hat{H} |He_{or}[n]> =E_{[n]} |He_{or}[n]>$
and $\hat{H} |He_{pa}[n]> =E_{[m]} |He_{pa}[m]>$ the stationary states of the superposition must obey the relation
\begin{equation}
E_{[n,m]}=E_{[n]}=E_{[m]}
\label{eq:ene}
\end{equation}
In general, the energies of ortho and para states are different. But
still it can happen that although with different energies, the
states be so close to be considered as almost degenerate. As we
shall discuss later in this section these states will decay by
spontaneous emission. Then their energies will have an uncertainty
determined by their mean lifetimes. In this context the natural
criterion to consider two states as almost degenerate is that the
difference between their energies be smaller than the broadening of
the lines. As the mean lifetimes are of the order $10^{-9}s$, the
uncertainty of the energies are $\delta E \approx 10^{-6}eV$ (note
that this value is of the same order of the actual precision on the
measurement of the energy). To see this point in detail let us
analyze the experimental data. By the matter of concreteness we use
the data of the NIST \cite{nis}. It is simple to see that there are
some states for which the condition in Eq. (\ref{eq:ene}) is
fulfilled up to the broadening of the lines (and the experimental
error). For instance, in the configuration $1s4f$ with terms
$^3F^0$, $J=2$ for the ortho and $^1F^0$, $J=3$ for the para we have
an energy difference between both levels $\Delta E=9 \, 10^{-7} eV$,
which is below $\delta E$ (and the experimental error). Similarly,
we have the configuration $1s5f$ with terms $^3F^0$,$J=2$ and
$^1F^0$,$J=3$, or other terms obeying the relation $\Delta E \leq
\delta E$.

Other degeneracy has already been signaled in the literature for
this system. For $d3$ configurations the terms $2P$ and $2H$ are
degenerate due to the  symmetries existent in the problem.

We conclude that there are some energy levels of the ortho- and
para-Helium which can be considered as almost degenerate. However,
these states associated with almost degenerate levels would be of no
interest since they would be unstable because of spontaneous
emission. For instance, in the case of the state $1s4f$ the electron
in the state $4f$ would emit photons decaying consecutively to
states $3d$, $2p$ (final state in the ortho case) or $1s$ (final
state for the para configuration). In the absence of stationary
states of interest the most relevant states of the system would be
the metastable ones. We consider then in next section.

We must signal here that $F$ levels are involved in the only (up to
our knowledge) previous consideration of mixed multiplicity states
(see \cite{JPB} and references therein).
They are states that cannot
be considered as pure singlets or triplets ones but rather as
mixtures of them.
They are related to the F-cascade model, in
which it is assumed that in a collision process the excitation
energy is transferred from resonance ($n^1P$)-levels to mixed
$F$-levels: $He(1^1S)+He(n^1P) \rightarrow He(n^{mix}F)+ He(1^1S)$,
where $n^{mix}F$ represents a superposition of $n^1F$ and $n^3F$
states.

\section{Metastable states}

In the Helium atom the $1s2s$ state is
metastable for both para- (lifetime of $19.7 ms$) and ortho-type
(lifetime around $10^8 s$) configurations. In the first case the
decaying occurs through a two-photon electric dipole transition and
in the second via relativistic and spin-orbit interactions. Consequently, we expect their
superposition also to be a metastable state. We
shall explore its properties.

First of all, we note that the energies of the two configurations are
different, $E(1s2s, or)=E_{or} \neq E_{pa}=E(1s2s, pa)$. Then
in addition to the superposition of the spins we must have a small
indetermination in the energy of the state (initially carried by
the incident electron), giving rise also to a superposition of the
energy states. Both superpositions are compatible because energy and
spin are compatible variables, not affected by complementarity
relations.

Next we consider the lifetime of the state. As usual, the lifetime
of a state is evaluated as the inverse of its decay rate, $\tau = 1/
\Gamma $. If we denote by $|\phi _{or}>=|1s2s, or>$, $|\phi
_{pa}>=|1s2s, pa>$ and $|\psi >=|1s1s, pa>$ (the ground state, at
which both decay) the decay rate is given by
\begin{eqnarray}
\Gamma = |<\psi | \hat{H}_*| \alpha \phi _{or}+ \beta \phi _{pa}>|^2 =|\alpha M_{or} e^{-iE_{or}t/\hbar }+ \alpha M_{pa} e^{-iE_{pa}t/\hbar } |^2 = \nonumber \\
|\alpha |^2 \Gamma _{or} + |\beta |^2 \Gamma _{pa} + 2 Re(\alpha ^* \beta M_{or}^* M_{pa} e^{-i(E_{pa}-E_{or})t/\hbar })
\end{eqnarray}
where $M_i= <\psi | \hat{H}_*|\phi _i>$, $i=or,pa$, are the matrix
elements giving the transition rates, $\Gamma _i =|M_i|^2$. Note
that we use $\hat{H}_*$ instead of $\hat{H}$ to remark that
now relativistic and spin-orbit interactions are included.

$\Gamma $ is the instantaneous value of the decay rate. However, it
is unobservable. We must consider the averaged value
$\overline{\Gamma }$ over the time scale $T$ characteristic of the
observations, $ \overline{\Gamma } =1/T \int _0^T \Gamma dt $.
Assuming by simplicity the coefficients $\alpha $ and
$\beta $ and the matrix elements to be real we have for the interference term
$ \Gamma _{\alpha \beta }/T \int _0^T cos (\omega t) dt = \Gamma _{\alpha
\beta } sin(\omega T)/\omega T$ where $\Gamma _{\alpha \beta }=2\alpha \beta M_{or} M_{pa}$ and
$\omega = (E_{pa}-E_{or})/\hbar $. As $\omega \approx 10^{15}
s^{-1}$ and $|sin(\omega T)| \leq 1$ we have that $sin(\omega
T)/\omega T \approx 0$ for any $T$ much larger than $10^{-15}s$.
Consequently, for any realistic $T$ the interference term can be
neglected in the averaged expressions.
Finally, we can write
\begin{equation}
\tau =\frac{1}{|\alpha |^2 \Gamma _{or} + |\beta |^2 \Gamma _{pa} }
\end{equation}
The lifetime of the superposition state ranges between $\tau
_{or}=1/\Gamma _{or}$ and $\tau _{pa}=1/\Gamma _{pa}$. Varying the
coefficients $\alpha $ and $\beta $ we can obtain all the values of
lifetimes in that range. For any value of the coefficients we have a
metastable state.

Although the interference term has not influence on the (mean)
lifetimes its effects
lead to high frequency oscillations
of the decay rate, in a similar way to the quantum beats present in
energy-type superposition states. To emphasize this point we follow
the approach of Ref. \cite{Sil} expressing the coefficients $\alpha
M_{or}$ as $A_{or}e^{-\Gamma _{or}t}$,..., i. e., taking into
account explicitly the dependence of the coefficients on the decay
rate. Hence, the probability for the transition at time $t$ (when
all the amplitudes are assumed to be real valued) is $A_{or} ^2
e^{-2\Gamma _{or} t} + A_{pa} ^2 e^{-2 \Gamma _{pa}t} + 2 A_{or}
A_{pa} e^{-(\Gamma _{or} + \Gamma _{pa})t} cos (\omega t)$, showing
clearly the existence of oscillations around the mean values. These
oscillations, being their amplitude and frequency of the same order
of magnitude than those associated with energy-type superpositions,
are in principle experimentally observable. Moreover, varying the
parameters $\alpha $ and $\beta $ we could modulate the amplitude of
the oscillations.

Another experimental way to test the existence of the superpositions
would be the measurement of the variations of the decaying rate. We
proceed in the standard way, i. e., by counting the number of decays
in a given time interval of observation. From an experimental point
of view the way of measuring the decays is to count the photons
emitted during that interval. As different atoms emit their
radiation independently this photon source is of chaotic type. As it
is well known \cite{Lou} the distribution of photocounts is of
Poisson type, $P_n (T)=<n>^n e^{-<n>}/n!$, provided that the time of
observation $T$ is much longer than the coherence time of the light
(if not the distribution would be super-Poisson). In the above
relation $<n>$ denotes the mean number of photocounts, which in the
semiclassical approximation (for chaotic light the semiclassical and
fully quantum approaches give the same result \cite{Lou}) is given
by the expression $<n>=\xi \overline{I}T$, with $\xi$ the efficiency
of the detectors and $\overline{I}=\overline{I}(t)$ the
cycle-averaged intensity of the light. The intensity is given by the
number of photons emitted at $t$. By definition, this number is
proportional to the number of atoms that decay at $t$, which is
$\Gamma $, and $I(t) \sim \Gamma (t)$. Finally, we must average over
the time of observation, which must be much longer than the
coherence time of the light. Denoting by $\overline{\Gamma}$ this
average over $T$ (which according to our previous results is
time-independent for any realistic choice of $T$) the mean number of
photocounts is $<n>=\overline{\Gamma}T$, where the efficiency factor
has been absorbed in $\overline{\Gamma}$ by simplicity in the
notation. The Poisson distribution can be expressed as
\begin{equation}
P_n(T)= \frac{(\overline{\Gamma } T)^n}{n!} exp(-\overline{\Gamma
}T)
\label{eq:Poi}
\end{equation}
Using the relation $\overline{\Gamma }=|\alpha |^2 \Gamma _{or} +
|\beta |^2 \Gamma _{pa}$ and the well-known expression $(x+y)^n =
\sum  x^{n_x} y^{n_y}n!/n_x ! n_y !$, where the summation extends to
all the non-negative integers obeying the relation $n_x + n_y = n$,
we have,
\begin{equation}
P_n (T)= \sum _{n_{or}+ n_{pa}=n} {\cal P}_{n_{or}} (T) {\cal P}_{n_{pa}} (T)
\label{eq:nin}
\end{equation}
where
\begin{equation}
{\cal P} _n(T)= \frac{(|\alpha |^2 \Gamma _{or} T)^{n_{or}}}{n_{or}!} exp(-|\alpha |^2 \Gamma _{or} T)
\end{equation}
that is, the same (\ref{eq:Poi}) distribution, but with the decay
rate replaced by $|\alpha |^2 \Gamma _{or}$, a weighted decay rate.
Therefore, the detection distributions show a characteristic
dependence on $\alpha $ and $\beta $ that could be tested
experimentally. Moreover, the distribution (\ref{eq:nin}) differs
from that expected for a mixture of Helium atoms prepared in
($1s2s$) ortho and para states with weights $|\alpha |^2$ and
$|\beta |^2$, which as it is simple to see is given by the
expression $ P_n^{mix} (T)= \sum _{n_{or}+ n_{pa}=n} |\alpha |^2
P_{n_{or}}^{or} (T) |\beta |^2 P_{n_{pa}}^{pa} (T)$, with $
P_{n_{or}}^{or}$ and $ P_{n_{pa}}^{pa}$ the usual Poisson
distributions with $\Gamma _{or}$ and $\Gamma _{pa}$.

We conclude that, in principle, some peculiar characteristics of the
metastable $1s2s$ superposition state can be observed
experimentally.

\section{Acknowledgments}
We acknowledge partial support from MEC (FIS2006-04151, Consolider
Program SAUUL, CSD2007-00013)

\end{document}